\documentclass[runningheads]{llncs}
\usepackage{graphicx}
\usepackage{booktabs}
\usepackage{subcaption}
\usepackage{xcolor}

\renewcommand{\vec}[1]{\underline{#1}}

\newcommand{\avg}[1]{\langle{#1}\rangle}
\newcommand{\flu}[1]{{#1}'}

\newcommand{\picbox}[1]{{#1}}

\graphicspath{{./figures/}}

\begin{document}
%
%\title{Contribution Title\thanks{Supported by organization x.}}
\title{Towards prediction of turbulent flows at high Reynolds numbers using high performance computing data and deep learning}
\titlerunning{Towards prediction of turbulent flows using HPC data and DL}
% Authors ordered as in Computing Time Proposol -> Can be changed
%
\author{Mathis Bode\inst{1}\orcidID{0000-0001-9922-9742} \and
Michael Gauding\inst{2}\orcidID{0000-0003-0038-5249} \and
Jens Henrik G\"obbert\inst{3}\orcidID{0000-0002-3807-6137} \and
Baohao Liao\inst{1}\orcidID{0000-0001-8335-4573} \and
Jenia Jitsev\inst{3}\orcidID{0000-0002-1221-7851} \and
Heinz Pitsch\inst{1}\orcidID{0000-0001-5656-0961}}
\authorrunning{M. Bode et al.}
\institute{Institute for Combustion Technology, RWTH Aachen University, Templergraben 64, 52062 Aachen, Germany \\
\email{\{m.bode,h.pitsch\}@itv.rwth-aachen.de} \and
CORIA -- CNRS UMR 6614, Saint Etienne du Rouvray, France\\
\email{michael.gauding@coria.fr} \and
J\"ulich Supercomputing Centre, FZ J\"ulich, Wilhelm-Johnen-Stra{\ss}e, 52425 J\"ulich, Germany \\
\email{\{j.goebbert,j.jitsev\}@fz-juelich.de}}
\maketitle              % typeset the header of the contribution
\begin{abstract}
  In this paper, deep learning (DL) methods are evaluated in the
  context of turbulent flows. Various generative adversarial networks
  (GANs) are discussed with respect to their suitability for
  understanding and modeling turbulence. Wasserstein GANs (WGANs) are
  then chosen to generate small-scale turbulence. Highly resolved
  direct numerical simulation (DNS) turbulent data is used for
  training the WGANs and the effect of network parameters, such as
  learning rate and loss function, is studied. Qualitatively good
  agreement between DNS input data and generated turbulent structures
  is shown. A quantitative statistical assessment of the predicted
  turbulent fields is performed.

\end{abstract}
\section{Introduction}
The turbulent motion of fluid flows is a complex, strongly non-linear, multi-scale phenomenon, which poses some of the most difficult and fundamental problems in classical physics. Turbulent flows are characterized by random spatio-temporal fluctuations over a wide range of scales. The general challenge of turbulence research is to predict the statistics of these fluctuating velocity and scalar fields. A precise prediction of these statistical properties of turbulence would be of practical importance for a wide field of applications ranging from geophysics to combustion science.

Research in the field of turbulence has mostly focused on a
statistical description in the sense of Kolmogorov's scaling
theory. The theory proposed by
Kolmogorov~\cite{kolmogorov1941,kolmogorov1941b} (known as K41 in
literature) hypothesizes that for sufficiently large Reynolds numbers,
small-scale motions are statistically independent from the large
scales. While the large scales depend on the boundary or initial
conditions, the smallest scales should be statistically universal and
feature certain symmetries that are recovered in a statistical
sense. Following Kolmogorov's theory, the small scales can be uniquely
described by simple parameters, such as the kinematic viscosity $\nu$
of the fluid and the mean dissipation rate $\avg{\varepsilon}$
(angular brackets denote ensemble-averaging). If the notion of
small-scale universality was strictly valid, then there would be
realistic hope for a statistical theory for turbulent flows. However,
numerous experimental and numerical studies have reported a
substantial deviation from Kolmogorov's classical K41
prediction~\cite{frisch1995,sreenivasan1996}, which is mostly
due to internal intermittency. The consequence of internal
intermittency is the break-down of small-scale universality, which
dramatically complicates theoretical approaches from first principles.

In this work, a novel research route based on the method of deep
learning (DL) is used to approach the challenge of turbulence
modeling. In recent years, DL was improved substantially and has
proven to be useful in a large variety of different fields, ranging
from computer science to life science. However, to the knowledge of
the authors, the application of DL to predict statistical behavior
of small-scale turbulence is still new and many related issues are still unsolved. As described and despite its stochastic
nature, turbulence exhibits certain coherent structures and
statistical symmetries that are traceable by deep learning
techniques. While analytical solutions exist for low-order correlation
functions, for higher orders there is no such tractable solution
available so far. Therefore, DL techniques are a promising approach to
predict statistics of small-scale turbulence and an attempt to
predict structures of turbulence is given here. Several DL networks
from literature are tested by training them with high-fidelity direct
numerical simulation (DNS) data of turbulence. The predicted
turbulence data is evaluated by qualitative and quantitative
comparisons with the original data and the statistics of the original
data, respectively. As one challenge in the application of DL is to
find optimal network architectures and hyperparameters, several
combinations were evaluated for this work.

The remainder of this paper is organized as follows. In
Sec.~\ref{sec:DLT}, future chances and challenges of DL in the context
of turbulence are summarized. Then, the used DNS data base is described in Sec.~\ref{sec:PW}. Section~\ref{sec:res}
presents results in terms of predicted turbulent structures and
discusses the sensitivity of the results with respect to network
parameters, such as learning rate and loss function. The paper finishes
with conclusions in Sec.~\ref{sec:con}.

\section{Deep Learning and Turbulence}
\label{sec:DLT}
The Reynolds number is the most important parameter for characterizing
turbulent flows.  It can be defined as the ratio of the size of the
large vortices to the size of the smallest vorticies and can be
understood as a measure for the scale
separation~\cite{pope2000turbulent}. Thus, it also plays a central
role in any attempt to model turbulence accurately and must be
considered in the DL network. As the application of DL for predicting
turbulence is new, the following steps need to be taken on the way to
find a turbulence model based on DL:
  \begin{enumerate}
  \item \textbf{A posteriori analysis with fixed Reynolds number:} The
    ability of certain network types to predict statistics of
    turbulence needs to be evaluated. Therefore, the Reynolds number
    should be fixed and unsupervised learning can be performed. The
    accuracy of the trained DL networks can be evaluated by comparison
    with the original data and corresponding statistics. The ability
    to predict statistics for a given Reynolds number is essential for
    an improved understanding of universality and intermittency as
    well as for the development of models.
  \item \textbf{A posteriori analysis with flexible Reynolds numbers:}
    As a next step, DL should be used with flexible Reynolds
    numbers. A combination of unsupervised and supervised DL can be
    employed to improve the understanding of the universality of
    turbulence. For example, it will be interesting to see whether a
    DL network, which was trained within a certain range of Reynolds
    numbers, is able to also predict turbulent structures for higher
    Reynolds numbers correctly.
  \item \textbf{A priori analysis:} Also, DL can be used to identify characteristic structures of the turbulent dissipation field by a pattern recognition technique.  Relevant quantities under consideration should be the moments of the dissipation $M_n$, the kinetic energy, and two-point correlation functions or structure functions of the velocity field.
  \item \textbf{Modeling of small-scale turbulence:} For many fields
    in turbulence research, small-scale quantities are not known,
    either due to modeling of the small scales in numerical
    approaches or due to lack of resolution in experimental
    techniques. For reduced order models of turbulence, it is required
    to be able to predict statistics of the dissipation without
    knowing the actual dissipation field. The goal is to develop a DL
    network that is able to predict statistics of the fine-scale
    motion by knowing the exact velocity field from DNS or a
    coarse-grained velocity field only. The reliability of the neural
    network can be statistically evaluated against data from DNS,
    keeping in mind that neural networks can fail under certain
    conditions.
  \end{enumerate}

\section{DNS Data Base}
\label{sec:PW}
The application of DL is only possible if a sufficiently large and
accurate data base exists. In recent years, a comprehensive data base
of DNSs has been created based on some
of the world's largest turbulence
simulations~\cite{gauding2013,gauding2015line,peters2016higher,gauding2017high}
using high performance computing (HPC). The data base contains
different flow setups, such as forced homogeneous isotropic turbulence
as well as free shear flows, and the data sets are freely available from the corresponding author upon request. DNS solves the governing equations of
turbulence (namely the Navier-Stokes equations) numerically for all
relevant scales, without relying on any turbulence models. Due to
internal intermittency, turbulent flows reveal a hierarchy of viscous
cut-off scales~\cite{boschung2016generalised} and the computation of
higher-order statistics of small-scale quantities requires a spatial
resolution that may be finer than the Kolmogorov length scale. DNS has
become an indispensable tool in the field of turbulence research as it
provides accurate access to three-dimensional (3-D) fields under controlled
conditions.

Characteristic properties of the DNSs are listed in
Table~\ref{tab:dns2}. $N$ denotes the number of grid points,
${\rm Re}_\lambda$ is the Reynolds number based on the Taylor
micro-scale, $\kappa_{\rm max}$ is the largest resolved wave-number, $\eta$ is
the Kolmogorov length scale, $\avg{\varepsilon}$ is the mean energy
dissipation. $M$ denotes the number of statistically independent boxes being available.
\begin{table}[h]
  \centering
  \caption{Characteristic parameters of the DNS data base. The DNS data base is used to train and test the neural network. }
  \label{tab:dns2}
  \begin{tabular}{l c c c c c c c c}\toprule
    & S &  R0      & R1 & R2 & R3 & R4 & R5 & R6 \\
    \midrule
    $ N$ & \mbox{ $1024 \times 512 \times 512 \;$ } & $512^3$    & $1024^3$    & $1024^3$    & $2048^3$    & $2048^3$    & $4096^3$ & $4096^3$\\
    $ {\rm Re}_\lambda$  & $\approx 50$   & 88        & 119   & 184  & 215 & 331   & 529 & 754\\
    $\nu$   & -             & 0.01    & 0.0055 & 0.0025 & 0.0019 & 0.0010 & 0.00048 & 0.00027 \\
    $\kappa_{\rm max} \eta$ & $> 2.5$  & 3.93      & 4.99    & 2.93    &  4.41    & 2.53
    & 2.95 & 1.60 \\
    $M$  & 1     & 189 & 62 & 61 & 10 & 10 & 10  & 11\\
    \bottomrule
  \end{tabular}
\end{table}

\begin{figure}[htbp]
 \centering
 \begin{subfigure}[c]{0.4\textwidth}
   \centering \includegraphics[width=\textwidth]{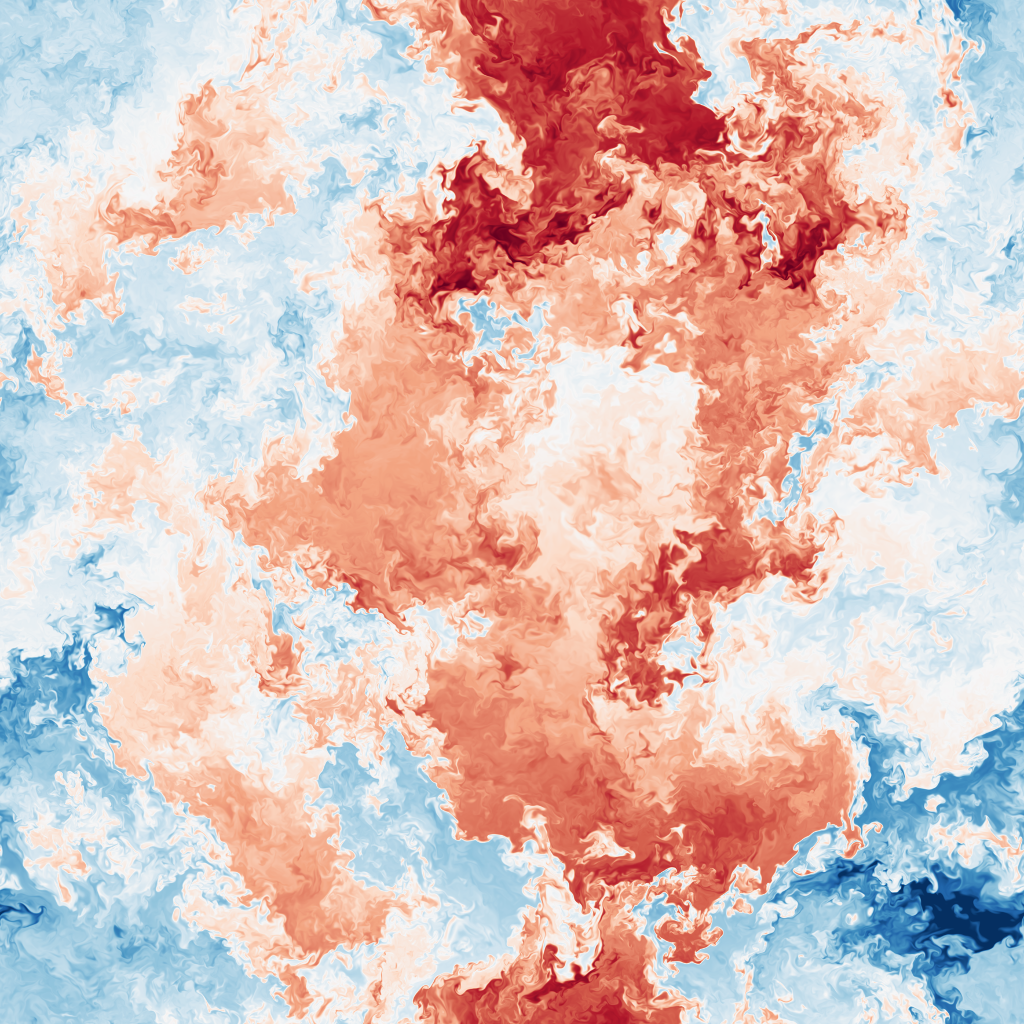}
   \subcaption{Scalar field}
 \end{subfigure}
 \begin{subfigure}[c]{0.4\textwidth}
 \centering
 \includegraphics[width=\textwidth]{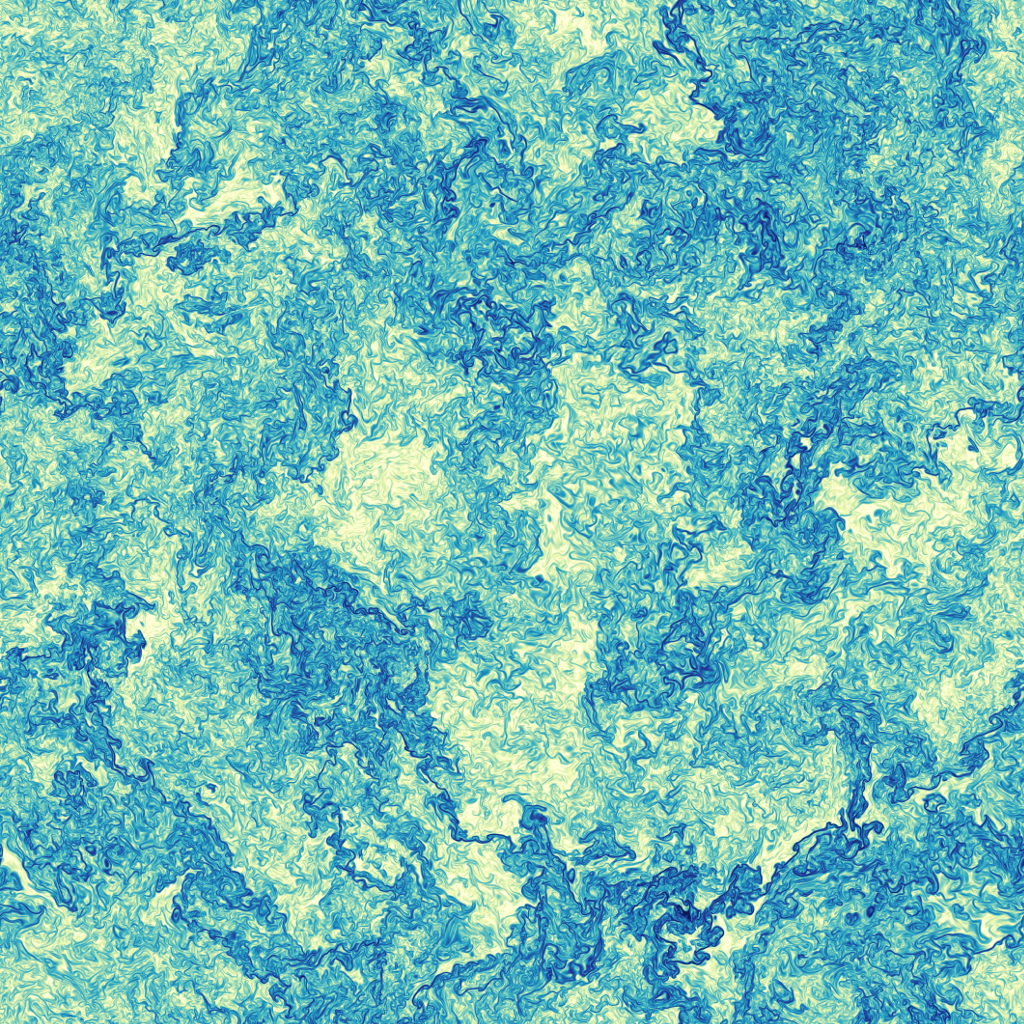}
 \subcaption{Scalar dissipation rate field}
 \end{subfigure}
 \caption{Visualization of the instantaneous fields of the scalar $\phi$ and the scalar dissipation rate $\chi$ for case R5.}
 \label{fig:example}
\end{figure}

Training of the DL is performed based on a passive scalar $\phi$ which
is transported by an advection-diffusion equation. The passive scalar
represents the dynamical motion of turbulence and its prediction by DL
is of fundamental relevance for the modeling of turbulence.
Figure~\ref{fig:example} displays a visualization of the instantaneous
scalar $\phi$ and the scalar dissipation rate $\chi$ for case R5. The
scalar dissipation rate is defined as
\begin{equation}
  \chi = 2D \left( \nabla \phi \right)^2
\end{equation}
and signifies the destruction of scalar fluctuations due to molecular
diffusivity $D$. The scalar field reveals distinct coherent
regions of roughly constant scalar values. The size of these regions
is of the order of the scalar integral length scale and they are
separated by sharp highly convoluted boundaries. At these boundaries,
the scalar dissipation rate attains large values. As a
consequence, the scalar dissipation rate is characterized by
filamented structures representing a high level of intermittency.

\section{Results}
\label{sec:res}
A first step towards the development of universal turbulent models based on DL is to study the ability of networks to generate small-scale turbulent structures. For that, slice-wise training of certain DL networks with the DNS data at fixed Reynolds number was performed. More precisely, the 3-D scalar fields $\phi$ of the DNS data are cut into 2-D slices, which are statistically similar. The distance between two adjacent slices is chosen large enough to ensure that structures are uncorrelated. The 2-D slices are used as input for the training of the network, which is able to reproduce these structures as output in the end. For this reproduction of turbulent structures, 100 random values following a normal Gaussian distribution with zero-mean and a standard deviation, which is linearly mapped between 0 and 1 by the local Reynolds number, are considered as input data. The implementation was done using Keras/TensorFlow/Horovod and the training was performed on JURECA, a supercomputer at JSC, FZ J\"ulich featuring two NVIDIA Tesla K80 GPUs with a dual-GPU design on each used computing node.

More precisely, for this work, generative adversarial networks
(GAN)~\cite{goodfellow2014}, which use an adversarial game between
generator and discriminator to optimize the network, were evaluated
regarding their suitability to generate small-scale
turbulence. Furthermore, Wasserstein GANs (WGANs)~\cite{arjovsky2017}
featuring advantages in terms of stability and good interpretability
of the learning curve as well as u-net~\cite{ronneberger2015},
consisting of a contracting path to capture context and a symmetric
expanding path that enables precise localization, were tried.

\begin{figure}[htbp]
 \centering
 \begin{subfigure}[c]{0.4\textwidth}
   \centering \includegraphics[width=\textwidth]{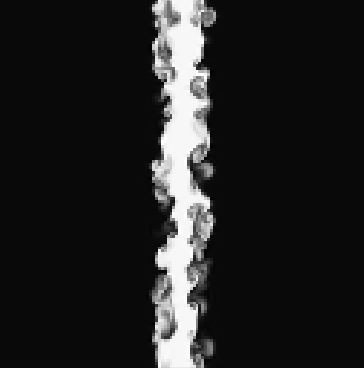}
   \subcaption{DNS (original)}
 \end{subfigure}
 \begin{subfigure}[c]{0.4\textwidth}
 \centering
 \includegraphics[width=\textwidth]{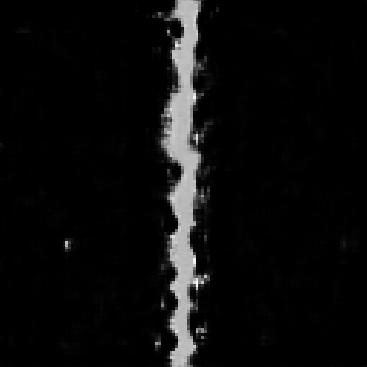}
 \subcaption{GAN (generated)}
 \end{subfigure}
 
  \begin{subfigure}[c]{0.4\textwidth}
   \centering \includegraphics[width=\textwidth]{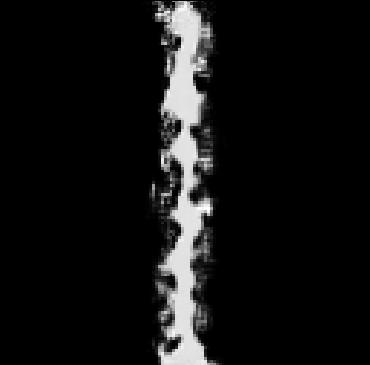}
   \subcaption{WGAN (generated)}
 \end{subfigure}
 \begin{subfigure}[c]{0.4\textwidth}
 \centering
 \includegraphics[width=\textwidth]{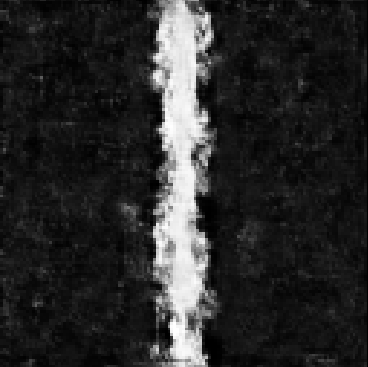}
 \subcaption{u-net (generated)}
 \end{subfigure}
 \caption{Comparison of an arbitrary DNS data slice of the passive
   scalar $\phi$ and comparison with data generated by various DL
   networks.}
 \label{fig:inout}
\end{figure}

A visualization of one arbitrary input slice cut from the DNS data and
one arbitrary output slice predicted by the DL networks for case S is
shown in Fig.~\ref{fig:inout}. The WGAN gives the best results and is
able to reproduce coherent motions and small-scale structures, which
are characteristic for fluid turbulence. GAN and u-net have problems
especially to reproduce regions without any fluctuations and always
feature some noise. Due to these results, the WGAN was tested in more
detail and quantitative results based on statistics in the original and predicted data are presented in the following.

As briefly mentioned in the beginning, one main challenge in the
context of DL is to find suitable network architectures and
hyperparameters resulting in an accurate solution. For the considered
turbulence, a WGAN made of four layers for the discriminator and three
layers for the generator gave good results. Each discriminator-layer
contained a convolution (Conv2D, kernel\_size=3, striding=2 or 1), an
activiation (LeakyReLU, alpha=0.2), and a dropout (Dropout). Partly,
zero-padding (ZeroPadding), batch normalization (BatchNormalization,
momentum=0.8), and flattening (Flatten) were employed. Each
generator-layer used a convolution (Conv2D, kernel\_size=4) and batch
normalization (BatchNormalization(momentum=0.8) in combination with
either tanh- or relu-activation.

\begin{figure}[htbp]
 \centering
  \begin{subfigure}[c]{0.8\textwidth}
 \centering \picbox{\input{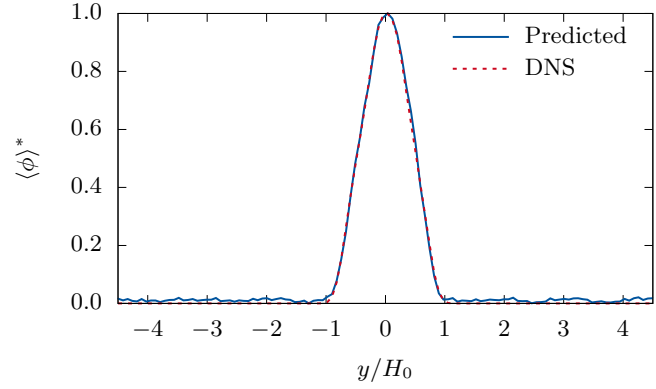}}
 \subcaption{$l_\mathrm{r}=0.00001$}
 \end{subfigure}
 \begin{subfigure}[c]{0.8\textwidth}
 \centering \picbox{\input{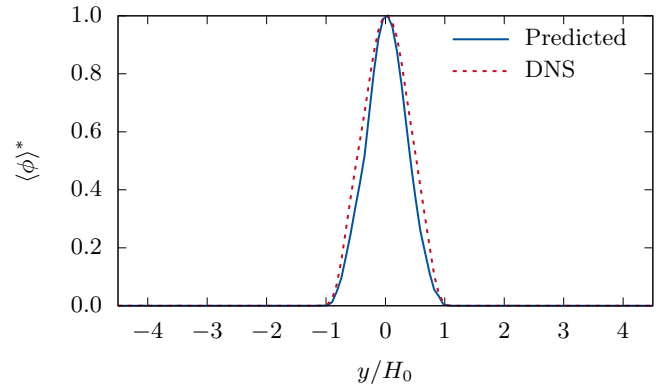}}
 \subcaption{$l_\mathrm{r}=0.0000005$}
 \end{subfigure}
 \begin{subfigure}[c]{0.8\textwidth}
 \centering \picbox{\input{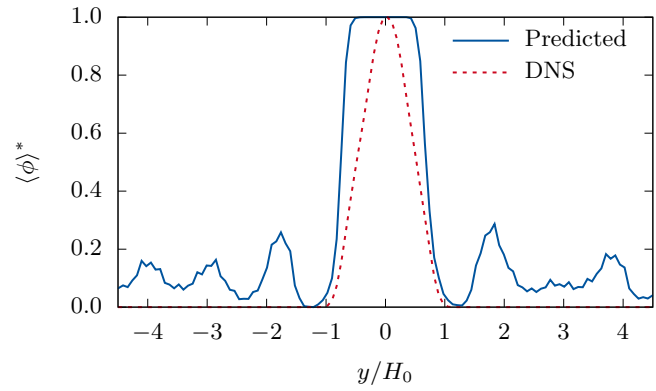}}
 \subcaption{$l_\mathrm{r}=0.00000001$}
  \end{subfigure}
 \caption{Comparison of the normalized scalar mean profile $\avg{\phi}^{*}$ for case S for different learning rates $l_\mathrm{r}$ plotted as function of the non-dimensional cross-stream direction $y/H_0$.}
 \label{fig:mean}
\end{figure}

A more rigorous quantitative analysis of the generated small-scale
turbulence is shown in Figs.~\ref{fig:mean} and \ref{fig:var}. It gives
the resulting normalized mean ($\avg{\phi}^{*}$) and variance profile ($\avg{\flu{\phi}^2}^{*}$) of the scalar for different learning
rates $l_\mathrm{r}$ as function of the non-dimensional cross-stream direction $y/H_0$. The maximum values are normalized to 1, $H_0$ is the initial jet width, and the asterisk indicates a normalized quantity. Furthermore, the scalar fluctuation is defined as $\flu{\phi}=\phi-\avg{\phi}$. It can be seen that the network is able to generate structures with the correct statistical properties as long as a proper learning rate is chosen. A
too small value for the learning rate leads to non-converged results,
while a too high value gives noise.
\begin{figure}[htbp]
 \centering
  \begin{subfigure}[c]{0.8\textwidth}
 \centering \picbox{\input{figures/v1_tex.tex}}
 \subcaption{$l_\mathrm{r}=0.00001$}
 \end{subfigure}
 \begin{subfigure}[c]{0.8\textwidth}
 \centering \picbox{\input{figures/v2_tex.tex}}
 \subcaption{$l_\mathrm{r}=0.0000005$}
 \end{subfigure}
 \begin{subfigure}[c]{0.8\textwidth}
 \centering \picbox{\input{figures/v3_tex.tex}}
 \subcaption{$l_\mathrm{r}=0.00000001$}
  \end{subfigure}
 \caption{Comparison of the normalized scalar variance $\avg{\flu{\phi}^2}^{*}$ for case S for different learning rates $l_\mathrm{r}$ as function of the non-dimensional cross-stream direction $y/H_0$.}
 \label{fig:var}
\end{figure} 

\begin{figure}[htbp]
 \centering \picbox{\input{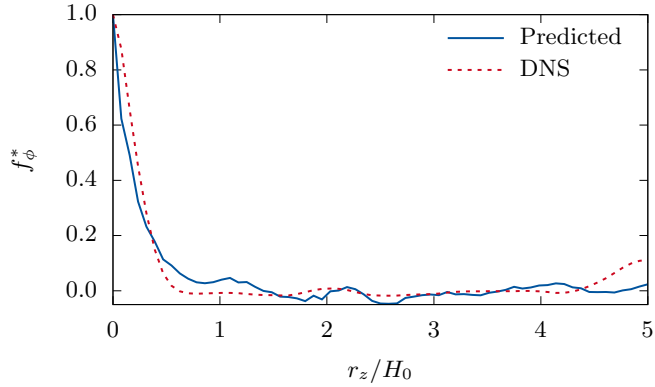}}
 \caption{Normalized scalar correlation function $f_{\phi}^{*}$ for case S
   as function of the non-dimensional separation distance in spanwise direction $r_z/H_0$. The correlation
   function is computed at the center-plane for learning rate $l_\mathrm{r}=0.0000005$.}
 \label{fig:corr}
\end{figure}

Finally, the DL approach is evaluated by means of two-point statistics. Turbulence is a non-local, multi-scale problem that is characterized by a transfer of turbulent energy from the large scales to the smaller scales, where the energy is dissipated due to molecular viscosity. Therefore, a two-point description of turbulence is customary as it captures both local and non-local phenomena. The two-point correlation function of the scalar field, defined as
\begin{equation}
  f_{\phi}(\vec{r};\vec{x}, t) = \avg{\flu{\phi}(\vec{x} + \vec{r}; t) \flu{\phi}(\vec{x}; t)}
\end{equation}
with underlines indicating vectors, is a statistical quantity of prime
importance and measures how the scalar field at the two independent
points $\vec{x} + \vec{r}$ and $\vec{x}$ is correlated. If the two
points are statistically independent then $f(\vec{r}; \vec{x},t)=0$.
The normalized correlation function of the scalar field, given by
\begin{equation}
  f_{\phi}^{*}(\vec{r};\vec{x}, t) =
  \frac
  {\avg{\flu{\phi}(\vec{x} + \vec{r}; t) \flu{\phi}(\vec{x}; t)}}
  {\avg{\flu{\phi}^2}},
\end{equation}
is shown in Fig.~\ref{fig:corr} for the DNS data and for the data obtained by DL evaluated at the center-plane in spanwise direction $z$ for a single timestep. A good agreement between the normalized correlation functions can be observed signifying that the DL approach is able to reproduce the local structure of turbulence with high accuracy. Different statistical quantities, such as the integral length scale $l_\mathrm{t}$, the scalar variance $\avg{\flu{\phi}^2}$, and the averaged scalar dissipation rate $\avg{\chi}$, can be computed from the correlation function. In other words, the ability to predict the correlation function correctly is a first step towards building a model for turbulence (cf.~\cite{de1938statistical}).

\section{Conclusions}
\label{sec:con}
In this work, DL is applied to turbulent fields obtained from DNS. The
effect of various network and training parameters, such as learning
rate and loss function, are discussed. Using WGANs, it was possible to
generate small-scale turbulence which features the same structures as
observed in DNS data. A comparison of the statistics evaluated on the
original and predicted data showed promising agreement. This is an
important first step towards the prediction of turbulent flows at various
Reynolds numbers. As a next step, the ability of the trained network to
generate turbulent structures for various given Reynolds numbers will
be evaluated.

\section*{Acknowledgment}
The authors gratefully acknowledge the computing time granted for the
project JHPC55 by the JARA-HPC Vergabegremium and provided on the
JARA-HPC Partition part of the supercomputer JURECA at
Forschungszentrum J\"ulich. Also, the computing time granted for the
projects HFG00/HFG02 on the supercomputer JUQUEEN~\cite{JSC2015} at
Forschungszentrum J\"ulich is acknowledged. MG acknowledges financial
support by Labex EMC3, under the grant VAVIDEN.

\bibliographystyle{splncs04}
\bibliography{literature}
%Author, F.: Article title. Journal \textbf{2}(5), 99--110 (2016)
%Author, F., Author, S.: Title of a proceedings paper. In: Editor,
%F., Editor, S. (eds.) CONFERENCE 2016, LNCS, vol. 9999, pp. 1--13.
%Springer, Heidelberg (2016). \doi{10.10007/1234567890}
%Author, F., Author, S., Author, T.: Book title. 2nd edn. Publisher,
%Author, A.-B.: Contribution title. In: 9th International Proceedings
%on Proceedings, pp. 1--2. Publisher, Location (2010)
%LNCS Homepage, \url{http://www.springer.com/lncs}. Last accessed 4
%\end{thebibliography}
\end{document}